\documentclass[twocolumn,showpacs,superscriptaddress,prb]{revtex4-1}
\usepackage{amsfonts}
\usepackage{amsmath}
\usepackage{amssymb}
\usepackage{graphicx}

\begin{document}

\title[The fine structure of magneto-oscillations]{The fine structure of microwave-induced magneto-oscillations in
photoconductivity of the two-dimensional electron system formed on a
liquid-helium surface}
\author{Yu.P. Monarkha}
\affiliation{Institute for Low Temperature Physics and Engineering, 47 Lenin Avenue, 61103
Kharkov, Ukraine}

\begin{abstract}
The influence of the inelastic nature of electron scattering by surface excitations
of liquid helium (ripplons) on the shape of magnetoconductivity oscillations induced
by resonance microwave (MW) excitation  is theoretically studied. The MW field provides
a substantial filling of the first excited surface subband which sparks off inter-subband
electron scattering by ripplons. This scattering is the origin of
magneto-oscillations in the momentum relaxation rate. The inelastic effect
becomes important when the energy of a ripplon involved compares with the collision
broadening of Landau levels. Usually, such a condition is realized only at sufficiently
high magnetic fields. On the contrary, the inelastic nature
of inter-subband scattering is shown to be more important in a lower
magnetic field range because of the new enhancement factor: the ratio of the inter-subband
transition frequency to the cyclotron frequency. This inelastic effect
affects strongly the shape of conductivity oscillations which acquires an additional
wavy feature (a mixture of splitting and inversion) in the vicinity of the
level-matching points where the above noted ratio is close to an integer.
\end{abstract}


%%73.40.-c    Electronic transport in interface structures
%%73.20.-r    Electron states at surfaces and interfaces
%%73.25.+i    Surface conductivity and carrier phenomena
%%78.70.Gq    Microwave and radio-frequency interactions
%%78.67.-n    Optical properties of low-dimensional, mesoscopic,
%%             and nanoscale materials and structures

%%\keywords{magnetooscillations, quantum magnetotransport, negative %%conductivity, zero resistance states}

\maketitle

\section{Introduction}

Microwave-induced magnetoconductivity oscillations, whose minima evolve into
novel zero-resistance states at high microwave (MW) power, were first
observed in degenerate semiconductor two-dimensional (2D) electron
systems~\cite{ZudSim-2001,ManSme-2002,ZudDu-2003}.
In these experiments, the magnetic field $\mathbf{B}$ was
directed normally to the electron layer, and the MW frequency $\omega $ was
substantially larger than the cyclotron frequency $\omega _{c}=eB/mc$. These
observations have attracted much theoretical interest and sparked invention
of a wide variety of theoretical mechanisms~\cite{DurSac-2003,DmiVav-2005,Mik-2011}
intended to explain remarkable 1/B-periodic oscillations and the appearance of
zero-resistance states.

Recently, similar 1/B-periodic magnetoconductivity oscillations were
observed in a particulary simple nondegenerate multi-subband 2D electron
liquid formed on the free surface of liquid $^{3}\mathrm{He}$\,~\cite{KonKon-2009,KonKon-2010}.
Their minima also evolve into zero-resistance states at high MW power.
Still, there is an important difference between results reported for surface
electrons (SEs) in liquid helium and those obtained for semiconductor systems.
In experiments~\cite{ZudSim-2001,ManSme-2002,ZudDu-2003}, $\omega $ is quite arbitrary: $\omega >\omega _{c}$.
To observe magneto-oscillations in the electron system formed on
the liquid-helium surface it is necessary that
the MW frequency $\omega $ be equal to the resonance frequency
$\omega _{2,1}=\left( \Delta _{2}-\Delta _{1}\right) /\hbar $ for
electron excitation to the next surface subband.
%%The important difference
%%of these experiments, as compared to those conducting on semiconductor systems,
%%%is that the MW frequency used in experiments on surface electrons in liquid helium
%%is fixed
%%two experimental observations is that the MW frequency used in experiments on
%%semiconductor systems is quite arbitrary ($\omega >\omega _{c}$%
%%), whereas for surface electrons on liquid helium it is fixed to .
The sequence $\Delta _{l}$ (here $%
l=1,2,...$) describes the energy spectrum of electron
subbands formed at the free surface of liquid helium because of the
interplay of the attractive image force and the potential barrier $V_{0}\sim
1\, \mathrm{eV}$ at the interface~\cite{ColCoh-1969,Shi-1970}.
In the limit of low
holding electric fields $E_{\bot }$ directed normally to the interface, $%
\Delta _{2,1}\equiv \Delta _{2}-\Delta _{1}\simeq 3.2\,\mathrm{K}$ for
liquid $^{3}\mathrm{He}$ and $\Delta _{2,1}\simeq 5.7\,\mathrm{K}$ for liquid
$^{4}\mathrm{He}$. Thus, in experiments on semiconductor systems, higher electron
subbands were not excited by the MW field, whereas in experiments with
SEs~\cite{KonKon-2009,KonKon-2010} there is a substantial fraction of electrons occupied
the first excited subband because of the MW resonance.

%The theoretical explanation of MW-resonance-induced magnetoconductivity
%oscillations observed for SEs on liquid helium was
%found~\cite{Mon-2011a,Mon-2011b} employing quasi-elastic approximation for electron
%scattering by ripplons and vapor atoms.
The theory explaining MW-resonance-induced magnetoconductivity oscillations
observed for SEs on liquid helium was worked out~\cite{Mon-2011a,Mon-2011b}
using the quasi-elastic approximation for electron
scattering by ripplons and vapor atoms.
This approximation assumes that
the energy exchange at a collision is much smaller than the
collision broadening of Landau levels.

The origin of magneto-oscillations and negative conductivity effects induced by
resonance MW excitation can be seen already from a simple analysis of the energy conservation
for usual inter-subband electron scattering (not involving photons) in the
presence of the uniform driving electric field $E_{\Vert }$. According to
this analysis, for the electron spectrum%
\begin{equation}
\varepsilon _{l,n,X}=\Delta _{l}+\hbar \omega _{c}\left( n+1/2\right)
+eE_{\Vert }X  \label{e1}
\end{equation}%
(here $X$ is the center coordinate of the cyclotron orbit), the decay of an excited subband $%
l\rightarrow l^{\prime }$ ($l>l^{\prime }$, $\Delta _{l,l^{\prime
}}=\Delta _{l}-\Delta _{l^{\prime }}>0$, and $n^{\prime }-n=m^{\ast }>0$)
results in displacements of the orbit center%
\begin{equation}
X^{\prime }-X=\frac{\hbar \omega _{c}}{eE_{\Vert }}\left( \frac{\Delta
_{l,l^{\prime }}}{\hbar \omega _{c}}-m^{\ast }\right)  \label{e2}
\end{equation}%
proportional to the quantity $\left( \Delta _{l,l^{\prime }}/\hbar \omega
_{c}-m^{\ast }\right) $. This quantity changes its sign at the
level-matching points $B_{m^{\ast }}$ defined by the condition $\Delta
_{l,l^{\prime }}/\hbar \omega _{c}=m^{\ast }$ (here $m^{\ast }$ is an
integer). At $B<$ $B_{m^{\ast }}$, the displacement $X^{\prime }-X>0$ and,
therefore, the decay of the excited subband is accompanied by electron
scattering against the driving force.

Somehow, Eq.~(\ref{e2}) resembles the displacement mechanism of the negative
conductivity effect discussed broadly for semiconductor systems~\cite{Ryz-1969}
where in a similar equation the photon quantum $\hbar \omega $ stands
instead of $\Delta _{l,l^{\prime }}$. In that theory, the sign-changing
correction to $\sigma _{xx}$ is due to radiation-induced disorder-assisted
current, and the   photon energy enters the energy conservation law. The
important point is that here photons are not involved in inter-subband
scattering directly and the condition $X^{\prime }-X>0$ found for decay
processes is not sufficient for obtaining negative corrections to the
momentum relaxation rate. Naturally, a reverse electron scattering from the
ground subband to the excited subband results in $X^{\prime }-X<0$, and the
negative conductivity effect is fully compensated if the electron system is
in equilibrium, and fractional occupancies of surface subbands obey the
condition $\bar{n}_{l}\equiv N_{l}/N_{e}=e^{-\Delta _{l,l^{\prime }}/T_{e}}%
\bar{n}_{l^{\prime }}$ (here $T_{e}$ is the electron temperature).

The important role of the resonant MW field is to provide an extra filling of the
excited subband to break the balance of inter-subband scattering
mentioned above.
For example, the decay of the first excited subband ($l=2$) to the ground
subband ($l=1$) caused by quasi-elastic scattering is possible only if $B$ is
close to the level-matching points $B_{m^{\ast }}$, because $\left\vert
X^{\prime }-X\right\vert $ is restricted by the magnetic length. If $B$ is
substantially away from these level-matching points, the quasi-elastic decay is
impossible and $\bar{n}_{2}\simeq \bar{n}_{1}\simeq 1/2$ because of MW
excitation. Under a weak driving field, the width of regions of the magnetic
field near $B_{m^{\ast }}$, where the excited subband decays
quasi-elastically, is determined by the collision broadening of the
corresponding Landau levels. Within these regions, the momentum relaxation
rate of SEs caused by inter-subband scattering~\cite{Mon-2011a,Mon-2011b}%
\begin{equation}
\nu _{\mathrm{inter}}\propto -\left( \frac{\omega _{2,1}}{\omega _{c}}%
-m^{\ast }\right) \left( \bar{n}_{2}-e^{-\Delta _{2,1}/T_{e}}\bar{n%
}_{1}\right) .  \label{e3}
\end{equation}%
Therefore, the condition $\bar{n}_{2}>e^{-\Delta _{2,1}/T_{e}}\bar{%
n}_{1}$ is crucial for the appearance of negative corrections to the
momentum relaxation rate $\nu $ and to magnetoconductivity $\sigma _{xx}$.
An increase in $\bar{n}_{2}$ caused by trivial heating of SEs obviously cannot
lead to the sign-changing term.

The above noted analysis assumes that the ripplon energy $\hbar \omega _{q}$
can be neglected as compares to typical electron energies. In the absence
of the magnetic field, this is the conventional approximation because $\hbar
\omega _{q}\ll T$ for ripplons involved in scattering events. Under
a magnetic field applied perpendicular to the electron layer, there is an
additional energy parameter which describes the width of the single-electron
density of states: the collision broadening of Landau levels $\Gamma _{n}$.
For SEs on liquid helium, Landau levels are extremely narrow: $%
\Gamma _{n}\ll T$. Wave vectors of ripplons involved in electron scattering
 $q\sim 1/L_{B}$ [here $L_{B}=\sqrt{\hbar c/eB}$ is
the magnetic length], and the energy exchange $\hbar \omega _{q}\propto q^{3/2}$
increases with $B$ faster than $\Gamma _{n}$ which is approximately
proportional to $\sqrt{B}$. Therefore, in a high magnetic field range,
depending on temperature ($\Gamma _{n}\propto \sqrt{T}$), $\hbar \omega _{q}$
becomes comparable with $\Gamma _{n}$, and electron scattering is
suppressed.

Experimental observation~\cite{MonItoShi-1997,MonKon-book} indicates
that the inelastic effect on the quantum magnetotransport becomes substantial
at $T\sim 0.1-0.2\, \mathrm{K}$ and $B>1\, \mathrm{T}$, and the suppression
of $\sigma _{xx}$ is the stronger the higher magnetic
field is applied. Simple estimates allow to assume that similar conditions
can be realized in an experiment on magneto-oscillations in
photoconductivity of SEs on liquid $^{4}\mathrm{He}$, because the
corresponding excitation frequency $\omega _{2,1}$ is high, and the
level-matching points $B_{m^{\ast }}>1\mathrm{T}$, if $m^{\ast }<4$.
%%Under the condition $\hbar \omega _{q}\gtrsim \Gamma _{n}$
In this case, the inelastic effect can cause
additional variations of the shape of magnetoconductivity oscillations
in the vicinity of the level-matching points $B_{m^{\ast }}$, which
could be used for experimental identification of the microscopic
mechanism of zero-resistance states and the resonant
photovoltaic effect discovered recently~\cite{KonCheKon-2012}.

In this work, we report the theory of magnetoconductivity oscillations
induced by resonance MW excitation which takes into account the inelastic
nature of the decay of excited subbands caused by electron-ripplon
interaction. We show that for inter-subband scattering the inelastic
effect displays differently as compared to the equilibrium magnetotransport in
a single subband. In our treatment, the maximum of the decay rate of an excited
subband is naturally split near
the level-matching points $B_{m^{\ast }}$  because of one-ripplon
creation and destruction processes. The unusual thing is that this inelastic
effect increases with $m^{\ast }$, and therefore extends itself into the
lower magnetic field range $B<1\mathrm{T}$.

We show also that the inelastic effect on the momentum relaxation rate
of the electron layer caused by inter-subband scattering cannot be reduced
to simple splitting similar to that of the decay rate.
In magnetoconductivity curves, this effect displays itself as
a combination of splitting and inversion. As a result, $\sigma _{xx}$
develops a new remarkable wavy feature in the vicinity of the level-matching
points. The influence of both mutual electron interaction and
electron heating on the new fine structure of MW-induced magneto-oscillations
is also analyzed.
%%This inelastic effect also enhances in the region of low magnetic
%%field $B<1\mathrm{T}$, which is contrary to the behavior of the inelastic
%%suppression of the equilibrium magnetoconductivity~\cite{MonItoShi-1997,MonKon-book}.

\section{Inelastic inter-subband scattering and momentum relaxation}

The most interesting features of MW-induced magnetoconductivity
oscillations such as zero-resistance states are
observed~\cite{KonKon-2009,KonKon-2010} in
the low temperature range ($T\sim 0.2\, \mathrm{K}$) where SEs are
predominantly scattered by capillary wave quanta (ripplons). Ripplons
represent a sort of 2D phonons with an unusual spectrum $\omega _{q}=\sqrt{%
\alpha /\rho }q^{3/2}$, where $\alpha $ and $\rho $ are the surface tension and
mass density of liquid helium, correspondingly. Therefore, the Hamiltonian of
electron-ripplon interaction is similar to
the Hamiltonian of electron-phonon interaction in solids%
\begin{equation}
H_{\mathrm{int}}^{(\mathrm{e-R})}=\sum_{e}\sum_{\mathbf{q}}U_{q}(z_{e})Q_{q}%
\left( b_{\mathbf{q}}+b_{-\mathbf{q}}^{\dag }\right) e^{i\mathbf{qr}_{e}},
\label{e4}
\end{equation}%
where $b_{\mathbf{q}}$ and $b_{\mathbf{q}}^{\dag }$ are destruction and
creation operators, $Q_{q}^{2}=\hbar q/2\rho \omega _{q}$, and $U_{q}(z_{e})$
is the electron-ripplon coupling whose properties and matrix elements $%
\left\langle l\right\vert U_{q}(z)\left\vert l^{\prime }\right\rangle \equiv
\left( U_{q}\right) _{l,l^{\prime }}$ are well defined in the
literature~\cite{MonKon-book,Mon-2012}.

To describe quantum magnetotransport of a 2D electron gas it is
conventional to use the self-consistent Born approximation (SCBA)
theory~\cite{AndUem-1974} or the linear response theory~\cite{KubMiyHas-1965}
with the proper approximation for the
electron density-of-states function. Unfortunately, these
approaches cannot be applied directly to the SE system under resonance MW
excitation. In these theories, conductivity is an equilibrium property
of the system, whereas here we need to describe conductivity of the
system which is far away from its equilibrium state. In our conductivity
treatment, we intend also to include strong Coulomb interaction between
electrons whose average potential energy can be much higher then the average
kinetic energy. For this purposes, it is necessary to use an extension of
the SCBA theory applicable for arbitrary subband occupancies $\bar{n}_{l}$
and $T_{e}\geq T$.

Firstly, we note that the well-known results of the SCBA theory
and Kubo equations for magnetoconductivity of the 2D electron gas can be
reproduced in a quite direct way by simple evaluation of the momentum gained
by scatterers~\cite{MonKon-book}, if scattering probabilities of the Born
approximation are taken in the proper form which includes the contribution from
high order terms (self-energy effects). This kind of probabilities
were actually given already in the Kubo theory~\cite{KubMiyHas-1965}. Here we express
these probabilities through a quite general correlation function of the
multi-subband 2D electron system which preserves basic equilibrium properties of the
in-plane motion and, at the same time, is independent of subband occupancies.

Consider the average probability of both intra and inter-subband scattering
($l\rightarrow l^{\prime }$) which is accompanied by the momentum exchange $%
\hbar \mathbf{q}$ caused by ripplon destruction $\bar{\nu}_{l,l^{\prime
}}^{\left( -\right) }\left( \mathbf{q}\right) $ and creation $\bar{\nu}%
_{l,l^{\prime }}^{\left( +\right) }\left( \mathbf{q}\right) $. Conventional
Born approximation yields%
\[
\bar{\nu}_{l,l^{\prime }}^{\left( -\right) }\left( \mathbf{q}\right) =\frac{%
2\pi \hbar }{A}u_{l,l^{\prime }}^{2}\langle \sum_{n^{\prime
},X^{\prime }}\left\vert \left( e^{i\mathbf{qr}_{e}}\right) _{l,X;l^{\prime
},X^{\prime }}\right\vert ^{2} \times
\]
\begin{equation}
\times \delta (\varepsilon _{n}-\varepsilon _{n%
\mathbf{^{\prime }}}+\Delta _{l,l^{\prime }}+\hbar \omega _{q}+eE_{\Vert
}X-eE_{\Vert }X^{\prime })\rangle _{\mathrm{in}},  \label{e5}
\end{equation}%
where $\varepsilon _{n}$ represents Landau levels, $\left\langle
...\right\rangle _{\mathrm{in}}$ means averaging over initial in-plane
states for the given surface subbad $l$, and we have introduced
\begin{equation}
\text{\ }u_{l,l^{\prime }}^{2}\left( x_{q}\right) =\frac{A}{\hbar ^{2}}%
N_{q}Q_{q}^{2}\left\vert \left( U_{q}\right) _{l,l^{\prime }}\right\vert
^{2}\simeq \frac{TL_{B}^{2}}{4\alpha \hbar ^{2}x_{q}}\left\vert \left(
U_{q}\right) _{l,l^{\prime }}\right\vert ^{2}  \label{e6}
\end{equation}%
as the function of the dimensionless parameter $x_{q}=q^{2}L_{B}^{2}/2$.
For $q\lesssim 1/L_{B}$, the distribution function of
ripplons $N_{q}\simeq T/\hbar \omega_{q}$.
In the following, the surface area $A$ will be set to unity.
It is well known that $\left\vert \left( e^{i\mathbf{qr}_{e}}\right)
_{l,X;l^{\prime },X^{\prime }}\right\vert ^{2}$ can be written as $%
J_{n,n^{\prime }}^{2}\left( q\right) \delta _{X^{\prime },X-q_{y}l_{B}^{2}}$%
, where $J_{n,n^{\prime }}^{2}\left( q\right) $ is a function of the
absolute value of the 2D wave-vector. The exact expression for $%
J_{n,n^{\prime }}^{2}\left( q\right) $ is given in the
literature (for recent examples, see~\cite{MonSokSmo-2010,Mon-2012}).

According the relationship $X^{\prime
}-X=-q_{y}L_{B}^{2}$ the quantity to be averaged in Eq.~(\ref{e5}) does not
depend on $X$, and, therefore, Eq.~(\ref{e5}) actually
contains the averaging over discrete Landau numbers $n$ only. It is natural
to assume that a weak dc driving field $E_{\Vert }$ does not change electron
distribution over Landau levels, and one can use the distribution function $%
e^{-\varepsilon _{n}/T_{e}}/Z_{\Vert }$ for the averaging operation.
This is quite clear in the absence of scatterers, because under the magnetic field
a driving electric field can be eliminated by a proper choice of the inertial
reference frame: $\mathbf{E}'=\mathbf{E}-\mathbf{B}\times \mathbf{V}/c\rightarrow 0$.
Moreover, if there is no a driving electric field in the laboratory frame,
it appears in any other inertial reference frame. At a low drift velocity
of the electron system, quasi-elastic scattering cannot change the population of Landau levels.
The above given statement is also verified by the comparison of the results obtained in
the treatment considered here with the well-known results of the conventional SCBA at zero
MW power.

Following the procedure described in the linear
response theory~\cite{KubMiyHas-1965},
and taking into account that $l_{B}^{2}eE_{\Vert }=\hbar V_{H}$ (here $%
V_{H}=cE_{\Vert }/B$ is the absolute value of the Hall velocity), Eq.~(\ref%
{e5}) can be transformed into the form containing level densities of the
initial and final states
\begin{equation}
\text{\ \ \ }\bar{\nu}_{l,l^{\prime }}^{\left( -\right) }\left( \mathbf{q}%
\right) =u_{l,l^{\prime }}^{2}S_{l,l^{\prime }}\left( q,\omega _{l,l^{\prime
}}+\omega _{q}+q_{y}V_{H}\right) ,\text{\ \ }  \label{e7}
\end{equation}%
where%
\[
S_{l,l^{\prime }}\left( q,\Omega \right) =\frac{2}{\pi \hbar Z_{\parallel }}%
\sum_{n,n^{\prime }}J_{n,n^{\prime }}^{2}(q) \times
\]
\begin{equation}
\times \int d\varepsilon
e^{-\varepsilon /T_{e}}\mathrm{Im}G_{l,n}\left( \varepsilon \right) \mathrm{Im}%
G_{l^{\prime },n^{\prime }}\left( \varepsilon +\hbar \Omega \right) .
\label{e8}
\end{equation}%
Here $G_{l,n}\left( \varepsilon \right) $ is the single-electron Green's
function of the corresponding subband whose imaginary part is a substitute
of $-\pi \hbar \delta \left( \varepsilon -\varepsilon _{n}\right) $. We
retain the index $l$ keeping in mind further broadening due to interaction
with scatterers because its strength is different for different surface subbands.
Similar equation can be found for ripplon creation processes:%
\begin{equation}
\bar{\nu}_{l,l^{\prime }}^{\left( +\right) }\left( \mathbf{q}\right)
=e^{\hbar \omega _{q}/T}u_{l,l^{\prime }}^{2}S_{l,l^{\prime }}\left(
q,\omega _{l,l^{\prime }}-\omega _{q}+q_{y}V_{H}\right) .  \label{e9}
\end{equation}%
At $l=l^{\prime }$ the function $S_{l,l^{\prime }}\left( q,\Omega \right) $
coincides with the dynamic structure factor (DSF) of a nondegenerate 2D
electron gas. It should be noted that the above given equations resemble
scattering cross-sections of thermal neutrons and X-rays in
solids~\cite{MarLov-1971}. Here scatterers (ripplons) play the role
of particle fluxes whereas the electron layer represents a target.

The self-energy effects (high order terms), which are very important for 2D
electron systems under a quantizing magnetic field, are taken into account
by inclusion of the collision broadening $\Gamma _{l,n}$ of Landau
levels of a given surface subband ($l$) according
to the cumulant approach~\cite{Ger-1976}:
\begin{equation}
-\mathrm{Im}G_{l,n}\left( \varepsilon \right) =\frac{\sqrt{2\pi }\hbar }{\Gamma
_{l,n}}\exp \left[ -\frac{2\left( \varepsilon -\varepsilon _{n}\right) ^{2}}{%
\Gamma _{l,n}^{2}}\right] .  \label{e10}
\end{equation}%
Thus, similarly to the Kubo presentation~\cite{KubMiyHas-1965},
the average probabilities
of electron scattering with the momentum exchange $\hbar \mathbf{q}$ are
expressed in terms of density-of-state functions of the initial and final
states broadened because of interaction with scatterers. For the Gaussian shape
of level densities, the integral entering the definition of $S_{l,l^{\prime
}}\left( q,\Omega \right) $ can be evaluated analytically~\cite{Mon-2011b}.
Moreover, this kind of a level density represents a good starting point for
obtaining an analytical form of $S_{l,l^{\prime
}}\left( q,\Omega \right) $ for the multi-subband 2D
Coulomb liquid~\cite{Mon-2012}.

The Eq.~(\ref{e8}) is a useful generalization of the DSF for the
multi-subband 2D electron system because it preserves the important property
of the equilibrium of the in-plane motion%
\begin{equation}
S_{l^{\prime },l}\left( q,-\Omega \right) =e^{-\hbar \Omega
/T_{e}}S_{l,l^{\prime }}\left( q,\Omega \right) ,  \label{e11}
\end{equation}%
and, at the same time, it does not depend on $\bar{n}_{l}$, which allows
to describe momentum relaxation for arbitrary subband occupancies.
The property of Eq.~(\ref{e11}) simplifies evaluations of the momentum
relaxation rate. For example,
using this property average probabilities for the reverse scattering
processes discussed in the Introduction can be transformed into the
same quantities of the direct processes:
\begin{equation}
\bar{\nu}_{l^{\prime },l}^{\left( +\right) }\left( \mathbf{q}\right)
=e^{-\Delta _{l,l^{\prime }}/T_{e}}e^{\hbar \omega _{q}\left(
1/T-1/T_{e}\right) }e^{\hbar q_{y}V_{H}/T_{e}}\bar{\nu}_{l,l^{\prime
}}^{\left( -\right) }\left( -\mathbf{q}\right) ,\text{ }  \label{e12}
\end{equation}%
\begin{equation}
\bar{\nu}_{l^{\prime },l}^{\left( -\right) }\left( \mathbf{q}\right)
=e^{-\Delta _{l,l^{\prime }}/T_{e}}e^{\hbar \omega _{q}\left(
1/T_{e}-1/T\right) }e^{\hbar q_{y}V_{H}/T_{e}}\bar{\nu}_{l,l^{\prime
}}^{\left( +\right) }\left( -\mathbf{q}\right) .  \label{e13}
\end{equation}%
We shall use these relationships in the following analysis.

The introduced above quantities $\bar{\nu}%
_{l,l^{\prime }}^{\left( \pm \right) }\left( \mathbf{q}\right) $ represent
useful assemblies to construct major relaxation rates of the multi-subband 2D
electron system under a quantizing magnetic field such as the decay rate of
an excited subband and the momentum relaxation rate due to inter-subband scattering.
It is important that they preserve peculiarities of quantum magnetotransport
in two-dimensions. They include
the self-energy effects eliminating magnetoconductivity singularities, the effect
of the driving electric field and, after the following generalization, can even
include strong Coulomb forces acting between electrons.

\subsection{The many-electron effect}

Even for the lowest electron areal density $n_{e}$ (about $1\times 10^{6}\,\mathrm{cm}^{-2}$)
used in the experiments on SEs in liquid helium~\cite{KonKon-2009,KonKon-2010},
Coulomb interaction between SEs cannot be neglected.
For example, at $T\simeq 0.2\, \mathrm{K}$ the
average interaction energy per an electron $U_{\mathrm{C}}$
is much larger than the average kinetic energy ($T$). The generalized DSF of
the multi-subband 2D electron system applicable for such conditions was
found in Ref.~\onlinecite{Mon-2012}:%
\begin{equation}
S_{l,l^{\prime }}\left( \Omega \right) =\frac{2\sqrt{\pi }\hbar }{%
Z_{\parallel }}\sum_{n,n^{\prime }}\frac{J_{n,n^{\prime }}^{2}}{\tilde{\Gamma%
}_{l,n;l^{\prime },n^{\prime }}}\exp \left[ -\frac{\varepsilon _{n}}{T_{e}}%
-D_{l,n;l^{\prime },n^{\prime }}\left( \Omega \right) \right] ,  \label{e14}
\end{equation}%
where%
\begin{equation}
D_{l,n;l^{\prime },n^{\prime }}=\frac{\hbar ^{2}\left( \Omega -m^{\ast
}\omega _{c}-\frac{\Gamma _{l,n}^{2}+x_{q}\Gamma _{C}^{2}}{4T_{e}\hbar }%
\right) ^{2}}{\tilde{\Gamma}_{l,n;l^{\prime },n^{\prime }}^{2}}-\frac{\Gamma
_{l,n}^{2}}{8T_{e}^{2}}\text{ },  \label{e15}
\end{equation}%
\begin{equation}
\tilde{\Gamma}_{l,n;l^{\prime },n^{\prime }}^{2}\left( x_{q}\right) =\frac{%
\Gamma _{l,n}^{2}+\Gamma _{l^{\prime },n^{\prime }}^{2}}{2}+x_{q}\Gamma
_{C}^{2}\text{ },  \label{e16}
\end{equation}%
$m^{\ast }=n^{\prime }-n$, $\Gamma _{C}=\sqrt{2}eE_{f}^{(0)}L_{B}$ and $%
E_{f}^{(0)}\simeq 3\sqrt{T_{e}}n_{e}^{3/4}$. The quantity $E_{f}^{(0)}$
represents the typical quasi-uniform electric field of other electrons
acting on a given electron because of thermal
fluctuations~\cite{DykKha-1979}. Here and
in some following equations we do not show explicitly the
dependence on $q$ of functions $S_{l,l^{\prime }}$, $%
J_{n,n^{\prime }}^{2}$ etc., in order to shorten lengthy equations. In the
limiting case $\Gamma _{C}\rightarrow 0$, Eq.~(\ref{e14}) transforms into the
generalized DSF of the multi-subband 2D system of noninteracting electrons.

The function $S_{l,l^{\prime }}\left( \Omega \right) $ has sharp maxima when
$\hbar \Omega $ is close to Landau excitation energies $\left( n^{\prime
}-n\right) \hbar \omega _{c}$. These maxima are broadened because of electron interaction
with scatterers and because of the fluctuational electric field $\mathbf{E}_{f}$%
. It is important that Coulomb broadening of the DSF $\sqrt{x_{q}}\Gamma
_{C} $ is not equivalent to the collision broadening because it depends on $%
q $ through the dimensionless parameter $x_{q}=q^{2}L_{B}^{2}/2$. The
fluctuational field does not broaden the single-electron level densities,
because, as noted above, it can be eliminated by a proper choice
of the inertial reference frame moving along the layer~\cite{MonKon-book}.

Small frequency shifts in the general expression for
$S_{l,l^{\prime }}\left( \Omega \right) $ play very important role: they provide
the basic property of Eq.~(\ref{e11}).
The small shift $\Gamma _{l,n}^{2}/4T_{e}\hbar $ can be neglected only for substantially
positive values of $\Omega $. Therefore, it is convenient to transform terms
containing $S_{l,l^{\prime }}\left( \Omega \right) $ with negative $\Omega $
into forms with positive $\Omega $ employing the relationship of
Eq.~(\ref{e11}). It worth noting also that the Coulomb shift in
the frequency argument of the DSF $x_{q}\Gamma_{C} ^{2}/4T_{e}\hbar $ increases with $x_{q}$
and $E_{f}^{(0)}$ faster than the Coulomb broadening $\sqrt{x_{q}}\Gamma
_{C} $ which curiously affects positions of magnetoconductivity
extremes~\cite{Mon-2012}. Therefore, we shall retain it in
$S_{l,l^{\prime }}\left( \Omega \right) $ even for substantially positive
values of the frequency argument.

\subsection{The decay rate of an excited subband}

The decay rate of an excited subband $l$ due to electron
scattering down to a lower subband $l^{\prime }<l$  is easily
expressed in terms of $\bar{\nu}_{l,l^{\prime }}^{\left( -\right) }\left(
\mathbf{q}\right) $ and $\bar{\nu}_{l,l^{\prime }}^{\left( +\right) }\left(
\mathbf{q}\right) $:%
\begin{equation}
\bar{\nu}_{l\rightarrow l^{\prime }}=\sum_{\mathbf{q}}\left[ \bar{\nu}%
_{l,l^{\prime }}^{\left( +\right) }\left( \mathbf{q}\right) +\bar{\nu}%
_{l,l^{\prime }}^{\left( -\right) }\left( \mathbf{q}\right) \right] .
\label{e17}
\end{equation}%
Here, we can neglect the small corrections $q_{y}V_{H}$ in the frequency
argument of the generalized DSF entering Eqs.~(\ref{e7}) and (\ref{e9}).
Then, using Eqs.~(\ref{e12}) and (\ref{e13}), one can see that for inelastic
scattering the detailed balancing $\bar{\nu}_{l^{\prime }\rightarrow
l}=e^{-\Delta _{l,l^{\prime }}/T_{e}}\bar{\nu}_{l\rightarrow l^{\prime }}$
is fulfilled only if the electron temperature coincides with the temperature
of the environment. Anyway, because of the condition $\hbar \omega _{q}\ll T$
discussed above, the detailed balancing is approximately valid even at high
electron temperatures.

When evaluating $\bar{\nu}_{l\rightarrow l^{\prime }}$ in the ultra-quantum
limit $\hbar \omega _{c}\gg T_{e}$ ($n=0$) one can use the approximate
expression for the generalized DSF,%
\begin{equation}
S_{l,l^{\prime }}\left( q,\omega _{l,l^{\prime }}\pm \omega _{q}\right)
\simeq 2\sqrt{\pi }\hbar \sum_{m=0}^{\infty }\frac{x_{q}^{m}e^{-x_{q}}}{m!%
\tilde{\Gamma}_{l,0;l^{\prime },m}}I_{l,l^{\prime };m}^{(\pm )}\left(
x_{q}\right) ,  \label{e18}
\end{equation}%
applicable for positive values of the frequency argument. Here we introduce
functions
\begin{equation}
I_{l,l^{\prime };m}^{(\pm )}\left( x_{q}\right) =\exp \left\{ -\left[
R_{l,l^{\prime };m}^{\left( \pm \right) }\left( x_{q}\right) \right]
^{2}\right\} ,  \label{e19}
\end{equation}%
and%
\begin{equation}
R_{l,l^{\prime };m}^{\left( \pm \right) }\left( x_{q}\right) =\frac{\left(
\hbar \omega _{l,l^{\prime }}\pm \hbar \omega _{q}-m\hbar \omega
_{c}-x_{q}\Gamma _{C}^{2}/4T_{e}\right) }{\tilde{\Gamma}_{l,0;l^{\prime },m}}%
,  \label{e20}
\end{equation}%
It should be noted that in Eqs.~(\ref{e19}) and (\ref{e20}) we have
omitted small frequency shifts of the order $\Gamma _{l,0}/4T_{e}$,
because the frequency argument of $S_{l,l^{\prime }}$
in Eq.~(\ref{e18}) is substantially positive
for decay processes.

Employing the above given notations $\bar{\nu}_{l\rightarrow l^{\prime }}$
can be transformed into
\[
\bar{\nu}_{l\rightarrow l^{\prime }}=\frac{T}{4\sqrt{\pi }\alpha
\hbar }\sum_{m=1}^{\infty }\frac{1}{m!}\int_{0}^{\infty }\frac{%
\left\vert \left( U_{q}\right) _{l,l^{\prime }}\right\vert ^{2}}{\tilde{%
\Gamma}_{l,0;l^{\prime },m}}x_{q}^{m-1}e^{-x_{q}}\times
\]
\begin{equation}
\times \left[ I_{l,l^{\prime
};m}^{(-)}\left( x_{q}\right) +I_{l,l^{\prime };m}^{(+)}\left( x_{q}\right) %
\right] dx_{q} .  \label{e21}
\end{equation}%
From this equation it is quite clear that the inelastic effect splits the
decay maximum of the elastic theory into two maxima when $\hbar \omega
_{q}=\hbar \sqrt{\alpha /\rho }2^{3/4}x_{q}^{3/4}/L_{B}^{3/2}$ becomes
comparable with the broadening $\tilde{\Gamma}_{l,0;l^{\prime },m}$%
. It is very important that the position of the maximum of the function $%
x_{q}^{m-1}e^{-x_{q}}$ entering the integrand of Eq.~(\ref{e21}) $\left(
x_{q}\right) _{\max }$ increases strongly with the level-matching number $m$%
, whereas $\Gamma _{l,0;l^{\prime },m}$ is nearly independent of $m$
if electron density is sufficiently low. This leads to unexpected enhancement
of the inelastic effect in the low magnetic field range
where $m^{*}=\mathrm{round}(\omega _{2,1}/\omega _{c})$ is
larger.

\begin{figure}[tbp]
\begin{center}
\includegraphics[width=11.cm]{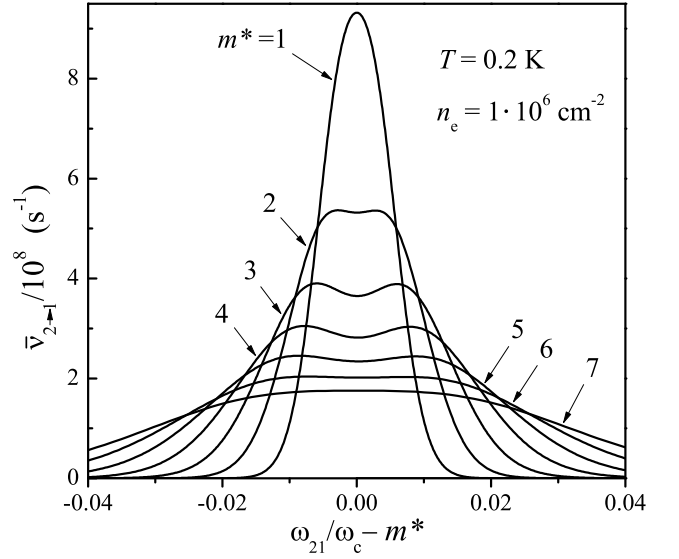}
\end{center}
\caption{The decay rate of the first excited subband $\bar{\nu}%
_{2\rightarrow 1}$ vs $\omega _{2,1}/\omega
_{c}\left( B\right) -m^{\ast }$ for different
level-matching numbers $m^{\ast }$.} \label{f1}
\end{figure}

The above stated is illustrated in Fig.~\ref{f1} where $\bar{\nu}%
_{2\rightarrow 1}$ is shown as the function of $\omega _{2,1}/\omega
_{c}\left( B\right) -m^{\ast }$ for $m^{\ast }=1,2,...,7$. Conditions of the
figure are chosen to be such ($n_{e}=1\times 10^{6}\, \mathrm{cm}^{-2}$ and $%
T=0.2\,\mathrm{K}$) that the splitting of the decay maximum is absent at $%
m^{\ast }=1$ ($B\rightarrow B_{1}$), however, it appears near some lower
level-matching points $B_{m^{\ast }}$. At the lowest $B_{m^{\ast }}$ ($m^{\ast }=7$), the
splitting disappears again because of the Coulombic correction $x_{q}\Gamma
_{C}^{2}$ to $\tilde{\Gamma}_{l,0;l^{\prime },m}^{2}$ which also increases
strongly with $m^{\ast }$.

\subsection{Subband occupancies}

Under the condition of the MW resonance $\omega =\omega _{2,1}$, the
stimulated photon absorption (emission) rate $r_{mw}=\Omega _{R}^{2}/2\gamma
_{mw}$, where $\gamma _{mw}$ is the half-width of the resonance, and $\Omega
_{R}=e\left\langle 2\right\vert z\left\vert 1\right\rangle E_{mw}/\hbar $ is
the Rabi frequency proportional to the amplitude of the MW field $E_{mw}$.
In dynamic equilibrium, the fractional occupancies $\bar{n}_{l}$ are found
from the time-independent rate equation. In the framework of the two-subband
model ($\bar{n}_{1}+\bar{n}_{2}=1$), the solution of the rate equation for
the relative occupancy has the following form~\cite{KonIssMon-2007}
\begin{equation}
\frac{\bar{n}_{2}}{\bar{n}_{1}}=\frac{r_{mw}+e^{-\Delta _{2,1}/T_{e}}\bar{\nu%
}_{2\rightarrow 1}}{r_{mw}+\bar{\nu}_{2\rightarrow 1}}.  \label{e22}
\end{equation}%
According to this equation the $1/B$-periodic dependence of the decay rate $%
\bar{\nu}_{2\rightarrow 1}$ with sharp maxima in the vicinity of the
level-matching points $B_{m^{\ast }}$ induces a $1/B$-periodic dependence
of the fractional occupancies $\bar{n}_{2}$ and $\bar{n}_{1}$.

\begin{figure}[tbp]
\begin{center}
\includegraphics[width=10.8cm]{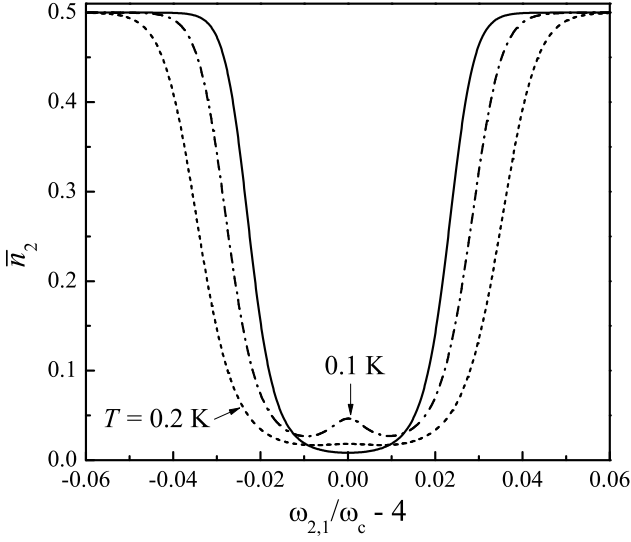}
\end{center}
\caption{The occupancy of the first excited subband vs $\omega _{2,1}/\omega
_{c}-4$ for $n_{e}=1\cdot 10^{6}\,\mathrm{cm}^{-2}$: elastic
treatment at $T=0.2\,\mathrm{K}$ (solid), inelastic theory for
$T=0.2\,\mathrm{K}$ (dashed) and $T=0.1\,\mathrm{K}$ (dash-dotted).
} \label{f2}
\end{figure}

In experiments~\cite{KonKon-2009,KonKon-2010}, the half-width
of the MW resonance was limited
by the inhomogeneous broadening ($\gamma _{mw}/\pi \simeq 0.3\, \mathrm{GHz}$).
We shall use this estimate in the following numerical evaluations.
For typical $\Omega _{R}/2\pi \simeq 15.9\,\mathrm{MHz}$, the results of
calculation of $\bar{n}_{2}$
are presented in Fig.~\ref{f2}. Variations of $\bar{n}_{2}$ are shown in the
vicinity of the level-matching point $B_{4}$ vs the parameter $\omega
_{2,1}/\omega _{c}-4$. Far away from the level-matching point, $\bar{\nu}%
_{2\rightarrow 1}$ is nearly zero and, therefore, $\bar{n}_{2}\rightarrow
\bar{n}_{1}\rightarrow 1/2$. In the vicinity of $B_{4}$, the occupancy $\bar{n%
}_{2}$ drops according to the sharp increase in the decay rate. The
inelastic effect broadens the $\bar{n}_{2}$ minima, and leads to small
local maxima at the level-matching points. The local peak at $\omega
_{2,1}/\omega _{c}=4$ becomes more pronounced with cooling, as shown in this
figure by the dash-dotted line.

For SEs on liquid helium at $E_{\bot }=0$, there is a spectrum crowding: $%
\Delta _{l}\rightarrow 0$ at $l\rightarrow \infty $. Therefore, at low
holding fields, there is a good chance to meet the MW resonance condition
for three surface subbands simultaneously: $\omega =\omega _{2,1}=\omega
_{k,2}$ where $k$ is substantially larger than $2$. In wide ranges between
the level-matching points $B_{m^{\ast }}$ introduced above for the first
excited subband, the decay rate $\bar{\nu}_{2\rightarrow 1}$ is very small,
and the occupancy of the third subband $\bar{n}_{k}$ can also satisfy the
condition $\bar{n}_{k}-e^{-\Delta _{k,1}/T_{e}}\bar{n}_{1}>0$ necessary for the
appearance of the sign-changing correction to $\sigma_{xx}$. Since $\omega
_{k,1}=2\omega _{2,1}$, the new level-matching condition $\omega
_{k,1}/\omega _{c}=m^{\ast }$ defined for the decay from $l=k$ to $l^{\prime
}=1$ can be rewritten as
\begin{equation}
\frac{\omega _{2,1}}{\omega _{c}}=m^{\ast }/2.  \label{e23}
\end{equation}%
Thus, oscillatory features of $\sigma _{xx}$ could appear also at fractional
values of the ratio $\omega /\omega _{c}$.

\subsection{Magnetoconductivity variations induced by the MW resonance}

Using the quantities $\bar{\nu}_{l,l^{\prime }}^{\left( +\right) }\left(
\mathbf{q}\right) $ and $\bar{\nu}_{l,l^{\prime }}^{\left( -\right) }\left(
\mathbf{q}\right) $ the frictional force $\mathbf{F}_{\mathrm{fric}}$
acting on the whole electron system can be evaluated directly: at first we
multiply these average probabilities by $N_{e}\bar{n}_{l}\hbar \mathbf{q}$,
and then perform summation over all $\mathbf{q}$ and the subband numbers $l$
and $l^{\prime }$. Since we intend to obtain the conductivity of interior
electrons ignoring edge effects, consider an uniform and infinite
electron layer. In this case, the kinetic friction $\mathbf{F}_{\mathrm{fric}%
}$ is antiparallel and proportional to the
current~\cite{PetSchLea-1994,MonTesWyd-2002,MonKon-book},
which can be written as $\mathbf{F}_{\mathrm{fric}}=-N_{e}m_{e}\nu _{\mathrm{%
eff}}\left\langle \mathbf{v}\right\rangle $, where $\left\langle \mathbf{v}%
\right\rangle $ is the average electron velocity. The proportionality factor
$\nu _{\mathrm{eff}}$ represents an effective collision frequency, because
balancing\ $\mathbf{F}_{\mathrm{fric}}$ and the average Lorentz force $%
\left\langle \mathbf{F}_{\mathrm{ext}}\right\rangle =-N_{e}e\mathbf{E}_{\Vert
}-N_{e}m_{e}\omega _{c}\left[ \left\langle \mathbf{v}\right\rangle \times
\mathbf{\hat{z}}\right] $ yields the conventional Drude form for the
conductivity tensor $\sigma _{i,k}$ with $\nu _{\mathrm{eff}}$ standing
instead of the semi-classical collision frequency. It should be emphasized
that here $\nu _{\mathrm{eff}}$ is not a semi-classical quantity because it
depends on $B$ and $n_{e}$.

The simplest way to obtain $\nu _{\mathrm{eff}}$ is to consider the
component $\left( \mathbf{F}_{\mathrm{fric}}\right) _{y}$ and to assume that
the magnetic field is strong enough ($\omega _{c}\gg \nu _{\mathrm{eff}}$): $%
\left\langle v_{y}\right\rangle \simeq -V_{H}$. Then, we can
represent $\nu _{\mathrm{eff}}$ as%
\begin{equation}
\nu _{\mathrm{eff}}=\frac{1}{m_{e}V_{H}}\sum_{l,l^{\prime }}\bar{n}_{l}\sum_{%
\mathbf{q}}\hbar q_{y}\left[ \bar{\nu}_{l,l^{\prime }}^{\left( +\right)
}\left( \mathbf{q}\right) +\bar{\nu}_{l,l^{\prime }}^{\left( -\right)
}\left( \mathbf{q}\right) \right] .  \label{e24}
\end{equation}%
According to (\ref{e7}) and (\ref{e9}), the right side of this relationship
is actually a nonlinear function of $V_{H}$.
Therefore, one have to expand it in $q_{y}V_{H}$. From the first glance at $%
S_{l,l^{\prime }}\left( \Omega \right) $ given in Eq.~(\ref{e14}) one may
conclude that $\hbar q_{y}V_{H}/\tilde{\Gamma%
}_{l,n;l^{\prime },n^{\prime }}$ is the main expansion parameter.
Still, an accurate analysis employing the
relationship of Eq.~(\ref{e11}) indicates that for intra-subband scattering at $T_{e}=T$
the actual expansion parameter equals $\hbar q_{y}V_{H}/T_{e}$.
Therefore, before proceeding with the expansion of the right side of Eq.~(\ref%
{e24}) we shall transform it into the form containing $S_{l,l^{\prime }}$ with
positive frequency arguments only.

For intra-subband scattering $\omega _{l,l^{\prime }}=0$, the basic property
of Eq.~(\ref{e11}) applied to $S_{l,l}\left( q,-\omega _{q}+q_{y}V_{H}\right)
$ yields
\[
\nu _{\mathrm{intra}}=\frac{1}{m_{e}V_{H}}\sum_{l}\bar{n}_{l}\sum_{\mathbf{q}%
}u_{l,l}^{2}\hbar q_{y}S_{l,l}\left( \omega _{q}+q_{y}V_{H}\right) \times
\]
\begin{equation}
\times \left[
1-e^{\hbar \omega _{q}\left( 1/T-1/T_{e}\right) }e^{-\hbar q_{y}V_{H}/T_{e}}%
\right] .  \label{e25}
\end{equation}%
Here we have changed the sign of the summation index $\mathbf{q}$ in the
term with $S_{l,l}\left( \omega _{q}-q_{y}V_{H}\right) $. At $T_{e}=T$ and
low $V_{H}$, the expression in the square brackets is proportional to $\hbar
q_{y}V_{H}/T_{e}$. Therefore, one can neglect $q_{y}V_{H}$ in the
frequency argument of the DSF. This confirms the above given statement that
for intra subband scattering $\hbar
q_{y}V_{H}/T_{e}$ is the main expansion parameter.

In the general case ($T_{e}\neq T$), one have to expand
the exponential function in $\hbar \omega _{q}\left( 1/T-1/T_{e}\right) $
and the DSF in $q_{y}V_{H}$ as well. This gives two terms ($\nu _{\mathrm{%
intra}}=\nu _{\mathrm{intra}}^{(0)}+\nu _{\mathrm{intra}}^{(1)}$):%
\begin{equation}
\nu _{\mathrm{intra}}^{(0)}=\frac{1}{m_{e}T_{e}}\sum_{l}\bar{n}_{l}\sum_{\mathbf{q}%
}u_{l,l}^{2}\hbar ^{2}q_{y}^{2}S_{l,l}\left( q,\omega _{q}\right) ,
\label{e26}
\end{equation}%
\begin{equation}
\nu _{\mathrm{intra}}^{(1)}=-\frac{T_{e}-T}{m_{e}TT_{e}}\sum_{l}\bar{n}_{l}\sum_{%
\mathbf{q}}u_{l,l}^{2}\hbar ^{2}q_{y}^{2}\omega _{q}S_{l,l}^{\prime }\left(
q,\omega _{q}\right) .  \label{e27}
\end{equation}%
Here and below $S_{l,l^{\prime }}^{\prime }=\partial S_{l,l^{\prime
}}/\partial \omega $. The term $\nu _{\mathrm{intra}}^{(0)}$ coincides with
the well-known result obtained previously for intra-subband
scattering~\cite{MonKon-book}. In the limit $n_{e}\rightarrow 0$, it transforms into the
result of the conventional SCBA theory~\cite{AndUem-1974}. The second term appears
only for $T_{e}\neq T$ when the scattering is substantially inelastic.
Therefore, it is not an equilibrium property of the system. In the absence
of the MW field, it can appear only as a nonlinear correction.

Consider now the contribution from inter-subband scattering. In Eq.~(\ref{e24}%
), one can transform terms with negative $\Omega $ ($l<l^{\prime }$) into the
forms with positive $\Omega $ using Eqs.~(\ref{e12}) and (\ref{e13}). Thus,
we have
\[
\nu _{\mathrm{inter}}=\frac{\hbar }{m_{e}V_{H}}\sum_{l>l^{\prime }}\sum_{%
\mathbf{q}}q_{y} \times
\]
\[
\huge{\{} \left[ \bar{n}_{l}-\bar{n}_{l^{\prime
}}e^{-\Delta _{l,l^{\prime }}/T_{e}}e^{\hbar \omega _{q}\left(
1/T_{e}-1/T\right) }e^{-\hbar q_{y}V_{H}/T_{e}}\right] \bar{\nu}%
_{l,l^{\prime }}^{\left( +\right) }\left( \mathbf{q}\right) +
\]%
\begin{equation}
+\left[ \bar{n}_{l}-\bar{n}_{l^{\prime }}e^{-\Delta _{l,l^{\prime
}}/T_{e}}e^{\hbar \omega _{q}\left( 1/T-1/T_{e}\right) }e^{-\hbar
q_{y}V_{H}/T_{e}}\right] \bar{\nu}_{l,l^{\prime }}^{\left( -\right) }\left(
\mathbf{q}\right) {\huge \}}. \label{e28}
\end{equation}%
The sign "-" of the second term in the square brackets appears because of
the change of the summation index $\mathbf{q\rightarrow -q}$ for terms
containing $\bar{\nu}_{l,l^{\prime }}^{\left( +\right) }\left( -\mathbf{q}%
\right) $ and $\bar{\nu}_{l,l^{\prime }}^{\left( -\right) }\left( -\mathbf{q}%
\right) $. Expanding this equation in $q_{y}V_{H}$ we find
that linear in $V_{H}$ terms of the square brackets yield a positive (normal)
contribution
\[
\nu _{\mathrm{inter}}^{(\mathrm{N})}=\frac{\hbar ^{2}}{m_{e}T_{e}}%
\sum_{l>l^{\prime }}\bar{n}_{l^{\prime }}e^{-\Delta _{l,l^{\prime
}}/T_{e}}\sum_{\mathbf{q}}u_{l,l^{\prime }}^{2}q_{y}^{2} \times
\]
\begin{equation}
\times {\LARGE \{}%
S_{l,l^{\prime }}\left( \omega _{l,l^{\prime }}-\omega _{q}\right)
+S_{l,l^{\prime }}\left( \omega _{l,l^{\prime }}+\omega _{q}\right) {\LARGE %
\}.}  \label{e29}
\end{equation}%
Here we used the condition $\hbar \omega _{q}\ll T$. If the electron system
is not heated high ($T_{e}\ll \Delta _{l,l^{\prime }}$), this contribution is
exponentially small.

Under MW excitation, the major contribution to $\nu _{\mathrm{inter}}$ comes
from the expansion of the DSF entering $\bar{\nu}_{l,l^{\prime }}^{\left(
+\right) }\left( \mathbf{q}\right) $ and $\bar{\nu}_{l,l^{\prime }}^{\left(
-\right) }\left( \mathbf{q}\right) $:
\[
\nu _{\mathrm{inter}}^{\left( \mathrm{A}\right) }=\frac{1}{m_{e}}%
\sum_{l>l^{\prime }}\sum_{\mathbf{q}}\hbar q_{y}^{2}u_{l,l^{\prime }}^{2}%
\times
\]
\[
{\LARGE \{}\left[ \bar{n}_{l}-\bar{n}_{l^{\prime }}e^{-\Delta _{l,l^{\prime
}}/T_{e}}e^{\hbar \omega _{q}\left( 1/T_{e}-1/T\right) }\right] \times
\]
\[
\times e^{\hbar
\omega _{q}/T}S_{l,l^{\prime }}^{\prime }\left( \omega _{l,l^{\prime
}}-\omega _{q}\right) +
\]%
\begin{equation}
+\left[ \bar{n}_{l}-\bar{n}_{l^{\prime }}e^{-\Delta _{l,l^{\prime
}}/T_{e}}e^{\hbar \omega _{q}\left( 1/T-1/T_{e}\right) }\right]
S_{l,l^{\prime }}^{\prime }\left( \omega _{l,l^{\prime }}+\omega _{q}\right)
{\LARGE \}}.  \label{e30}
\end{equation}%
These anomalous terms are proportional to the
derivative of $S_{l,l^{\prime }}\left( \Omega \right) $. The Eq.~(\ref{e30})
can be simplified considering the two-subband model with $T_{e}=T$ and $%
\hbar \omega _{q}/T\ll 1$. Then, we obtain%
\[
\nu _{\mathrm{inter}}^{\left( \mathrm{A}\right) }=\frac{\hbar }{m_{e}}\left(
\bar{n}_{2}-\bar{n}_{1}e^{-\Delta _{2,1}/T_{e}}\right) \times
\]
\begin{equation}
 \times \sum_{\mathbf{q}%
}u_{2,1}^{2}q_{y}^{2}\left[ S_{2,1}^{\prime }\left( \omega _{2,1}-\omega
_{q}\right) +S_{2,1}^{\prime }\left( \omega _{2,1}+\omega _{q}\right) \right]
.  \label{e31}
\end{equation}%
From this equation, one can see that the anomalous contribution $\nu _{\mathrm{%
inter}}^{\left( \mathrm{A}\right) }$ is proportional to $\bar{n}_{2}-\bar{n}%
_{1}e^{-\Delta _{2,1}/T_{e}}$ and to the sign-changing terms $\left( \omega
_{2,1}\pm \omega _{q}-m\omega _{c}\right) $, as expected from the
qualitative analysis given in the Introduction.

In the elastic theory, $%
\nu _{\mathrm{inter}}^{\left( \mathrm{A}\right) }$ changes its sign once in the
vicinity of each $B_{m^{\ast }}$. When the inelastic effect is substantial, the
expression in the square brackets is a derivative of the function which has
two maxima and one minima near each $B_{m^{\ast }}$. Therefore, in the
vicinity of a level-matching point, the anomalous contribution caused by inelastic
inter-subband scattering changes its sign three times.

It should be noted that here we use slightly different definitions of the
normal $\nu _{\mathrm{inter}}^{(\mathrm{N})}$ and
anomalous $\nu _{\mathrm{inter}}^{\left( \mathrm{A}\right) }$ contributions
than that given before~\cite{Mon-2011b,Mon-2012}.
%Previously, we defined this terms according to the origin of the expansion
%terms: the exponential function, or the DSF.  In this case,
%$\nu _{\mathrm{inter}}^{\left( \mathrm{A}\right) }$ has small terms
%which compensate terms of $\nu _{\mathrm{inter}}^{(\mathrm{N})}$, which
%are proportional to $\bar{n}_{2}$.
Now we apply labels N and A to the
corresponding expressions which are transformed into the form containing
the summation over $l>l^{\prime }$ only. For such definition,
$\nu _{\mathrm{inter}}^{(\mathrm{N}%
)}$ becomes substantially smaller,
and $\nu _{\mathrm{inter}}^{\left( \mathrm{A}\right) }$ does not contain small terms
which are not proportional to $S_{2,1}^{\prime }(\Omega )$ with $\Omega >0$.
%%can be reduced to the form similar to that of the normal contribution.
In the limiting case $\hbar \omega
_{q}\rightarrow 0$, the sum of $\nu _{\mathrm{inter}}^{\left( \mathrm{A}%
\right) }+\nu _{\mathrm{inter}}^{(\mathrm{N})}$ obviously coincides with
that found previously in the elastic treatment.

\section{Results and discussion}

\begin{figure}[tbp]
\begin{center}
\includegraphics[width=11.cm]{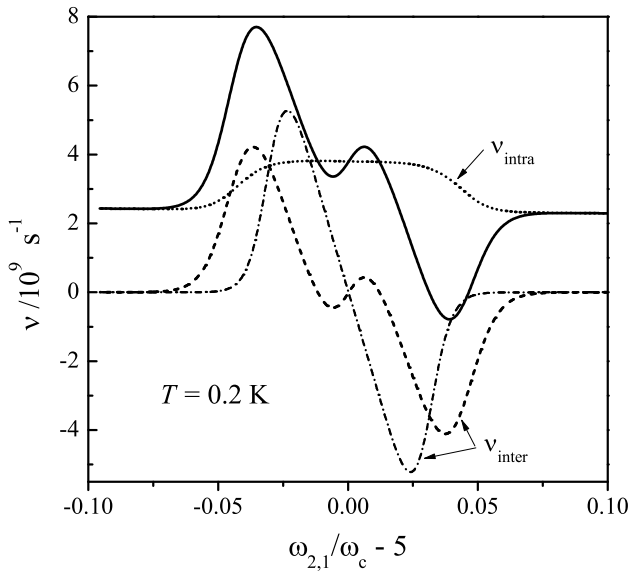}
\end{center}
\caption{Different contributions into the effective collision frequency
near the level-matching point $B_{5}$ vs the parameter
$\omega _{2,1}/\omega _{c}-5$ for $T_{e}=T=0.2\,\mathrm{K}$ and
$n_{e}=1\cdot 10^{6}\,\mathrm{cm}^{-2}$: inelastic treatment for
intra-subband scattering (dotted), inter-subband scattering
for elastic (dash-dotted) and inelastic (dashed) theories. The
total collision frequency is shown by the solid line.
} \label{f3}
\end{figure}

In the following evaluations, we shall consider strictly the approximation $%
T_{e}\simeq T$ and fix $\omega_{2,1}/2\pi$ to $140\,\mathrm{GHz}$ which
corresponds to $E_{\bot }=28\, \mathrm{V}/\mathrm{cm}$
for SEs on liquid $^{4}\mathrm{He}$.
The situation when $T_{e}$ differs substantially from $T$
will be analyzed only qualitatively. At typical temperatures of the
ripplon scattering regime $T\leq 0.3\, \mathrm{K}$, we can restrict ourselves to the
ultra-quantum limiting case ($\hbar \omega _{c}\gg T_{e}$). Then, Eq.~(\ref%
{e26}) can be represented as
\[
\nu _{\mathrm{intra}}^{(0)}=\frac{\omega _{c}T}{4\sqrt{\pi }\alpha T_{e}}%
\sum_{l}\bar{n}_{l} \times
\]
\begin{equation}
\times \int_{0}^{\infty }\frac{\left\vert \left( U_{q}\right)
_{l,l}\right\vert ^{2}}{\tilde{\Gamma}_{l,0;l,0}}\exp \left[ -x_{q}-\frac{%
\left( \hbar \omega _{q}\right) ^{2}}{\tilde{\Gamma}_{l,0;l,0}^{2}}\right]
dx_{q},  \label{e32}
\end{equation}%
This equations shows how the inelastic effect suppresses the contribution
from intra-subband scattering. Magneto-oscillations of $\nu _{\mathrm{intra}%
}^{(0)}$ are due to variations of $\bar{n}_{l}\left( B\right) $ discussed above
and $T_{e}\left( B\right) $ if electron heating becomes important. For $%
T_{e}=T$, the typical shape of these oscillations is shown in Fig.~\ref{f3} by
the dotted line. These evaluations have employed the $n_{l}\left( B\right) $
of the two-subband model. We do not show the corresponding line calculated for the
elastic approximation because it has a similar shape: just a narrower
and higher maximum.

In the same way, the nonequilibrium correction $\nu _{\mathrm{intra}%
}^{(1)}$ can be found as%
\[
\text{\ \ }\nu _{\mathrm{intra}}^{(1)}=\frac{T_{e}-T}{T_{e}}\frac{\omega _{c}%
}{2\sqrt{\pi }\alpha }\sum_{l}\bar{n}_{l} \times
\]
\begin{equation}
\times \int_{0}^{\infty }\frac{\left\vert
\left( U_{q}\right) _{l,l^{\prime }}\right\vert ^{2}}{\tilde{\Gamma}%
_{l,0;l,0}}\frac{\left( \hbar \omega _{q}\right) ^{2}}{\tilde{\Gamma}%
_{l,0;l,0}^{2}}\exp \left\{ -x_{q}-\frac{\left( \hbar \omega _{q}\right) ^{2}%
}{\tilde{\Gamma}_{l,0;l,0}^{2}}\right\} dx_{q}.  \label{e33}
\end{equation}%
This contribution becomes important when $T_{e}$ substantially exceeds $T$.
As compared to Eq.~(\ref{e32}), its integrand contains an additional parameter $%
\left( \hbar \omega _{q}/\tilde{\Gamma}_{l,0;l,0}\right) ^{2}$.

According to Eq.~(\ref{e32}) with an increase in $T_{e}$
the contribution $\nu _{\mathrm{intra}}^{(0)}$
decreases as $1/T_{e}$ if the Coulomb broadening $\Gamma _{C}$ can be
neglected. For a finite electron density, the Coulombic effect eventually changes
this dependence into $1/T_{e}^{3/2}$. On the contrary, $\nu _{\mathrm{intra}%
}^{(1)}$ increases first with $T_{e}$ approaching a saturation. Then, the
Coulombic effect leads to a decrease with the similar dependence $%
1/T_{e}^{3/2}$. Magneto-oscillations of $\nu _{\mathrm{intra}}^{(1)}$
crucially depend on electron temperature oscillations. Still,
the calculation of $T_{e}\left( B\right) $ requires the knowledge of the
electron energy relaxation rate which will not be discussed in the present
work.

As noted above, the contribution $\nu _{\mathrm{inter}}^{(\mathrm{N})}$ is
exponentially small, and it can be neglected in calculations
with $T_{e}\simeq T$. In the ultra-quantum limit, the contribution $\nu _{%
\mathrm{inter}}^{\left( \mathrm{A}\right) }$ can be transformed into

\[
\nu _{\mathrm{inter}}^{\left( \mathrm{A}\right) }=-\frac{\omega _{c}T}{2%
\sqrt{\pi }\alpha } \sum_{l>l^{\prime }}\left( \bar{n}_{l}-\bar{n}_{l^{\prime
}}e^{-\Delta _{l,l^{\prime }}/T_{e}}\right) \times
\]%
\[
\times \sum_{m=0}^{\infty }\frac{1}{m!}%
\int _{0}^{\infty }dx_{q} \frac{\left\vert \left( U_{q}\right) _{l,l^{\prime }}\right\vert
^{2}}{\tilde{\Gamma}_{l,0;l^{\prime },m}^{2}}x_{q}^{m}e^{-x_{q}}\times
\]
\begin{equation}
\times \left[ R_{l,l^{\prime };m}^{\left( +\right) }\left( x_{q}\right)
I_{l,l^{\prime };m}^{(+)}\left( x_{q}\right) +R_{l,l^{\prime };m}^{\left(
-\right) }\left( x_{q}\right) I_{l,l^{\prime };m}^{(-)}\left( x_{q}\right) %
\right] .  \label{e34}
\end{equation}%
This equation shows the way how the inelastic effect affects
magneto-oscillations of $\nu _{\mathrm{inter}}$.
Firstly, we note that the
integrand of Eq.~(\ref{e34}) contains the factor $x_{q}^{m}e^{-x_{q}}$ which
enhances the inelastic effect ($\omega _{q}\propto x_{q}^{3/4}$) in the low
field range where the level-matching numbers $m^{\ast }$ are larger. It
enhances also the Coulombic correction to the broadening parameter $\tilde{%
\Gamma}_{l,0;l^{\prime },m}$. Therefore, to observe the inelastic effect
on magneto-oscillations of $\sigma _{xx}$ electron densities must be sufficiently low.

The results of numerical evaluations of $\nu _{\mathrm{inter}}^{\left(
\mathrm{A}\right) }$ are given in Fig.~\ref{f3} for $B\approx B_{5}$.
Variations of $\nu _{\mathrm{inter}}^{\left( \mathrm{A}\right) }$ obtained
in the elastic approximation are shown by the dash-dotted line. It changes
sign only once. The dashed line calculated according to the inelastic
theory changes its sign three times. Remarkably, at $B=B_{5}$ the slope of
this line is changed to the opposite, as compared to the result of the
elastic theory. Thus, we have a fine oscillatory structure in the vicinity
of the level-matching point caused by the inelastic effect. The total
collision frequency $\nu _{\mathrm{eff}}$ is shown by the solid line. For
the chosen excitation rate, $\nu _{\mathrm{eff}}$ acquires negative values
which leads to negative conductivity effects. The condition $\sigma_{xx}<0$
was previously shown~\cite{AndAle-2003} to be the origin
of zero-resistance states.

\begin{figure}[tbp]
\begin{center}
\includegraphics[width=10.8cm]{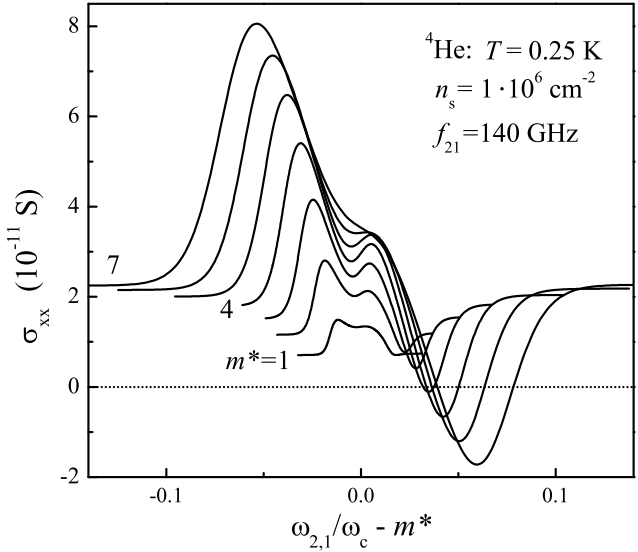}
\end{center}
\caption{Magnetoconductivity near the level-matching points $B_{m^{\ast }}$ vs
$\omega _{2,1}/\omega _{c}-m^{\ast }$ for $m^{\ast }=1,2,3,...,7 $.
} \label{f4}
\end{figure}

The evolution of the shape of magnetoconductivity variations near $B_{m^{\ast }}$ is
illustrated in Fig.~\ref{f4} for $m^{\ast }=1,2,...,7$ and $T_{e}=T$. The
inelastic effect displays itself as an additional wavy variation in the
vicinity of $\omega _{2,1}/\omega _{c}-m^{\ast }=0$. For $m^{\ast }=1$, the
inelastic effect is not strong and the corresponding line just shows an
additional plateau at the level-matching condition. The inelastic effect
becomes stronger for larger $m^{\ast }$ (lower magnetic fields): the amplitude of
new wavy variations of $\sigma _{xx}$ increases. Then, at $m^{\ast }=5$ the
amplitude of variations caused by the inelastic effect starts to decrease,
and at $m^{\ast }=6$ a new plateau appears at the level-matching point. This
reduction of the inelastic effect is caused by the corresponding increase in
the Coulomb broadening of the generalized DSF.

\begin{figure}[tbp]
\begin{center}
\includegraphics[width=11.cm]{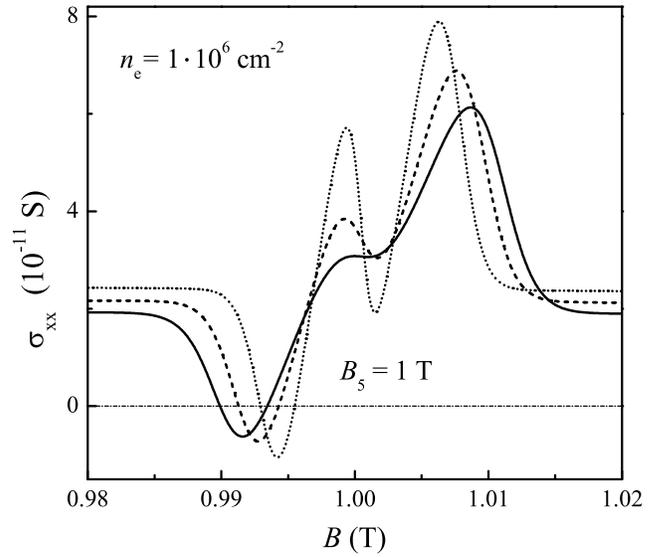}
\end{center}
\caption{Magnetoconductivity vs $B$ near the level-matching point $B_{5}$
for different temperatures: $0.3\,\mathrm{K}$ (solid), $0.2\,\mathrm{K}$
(dashed), and $0.1\,\mathrm{K}$ (dotted).
} \label{f5}
\end{figure}

The broadening of the DSF decreases with cooling $\tilde{\Gamma}%
_{l,0;l^{\prime },m}\propto \sqrt{T}$ if the holding field is low.
Therefore, the inelastic effect becomes more pronounced at lower
temperatures. This is illustrated in Fig.~\ref{f5} where $\sigma _{xx}$ is
plotted vs $B$ in the vicinity of $B_{5}=1\, \mathrm{T}$ for three different
temperatures. At $T=0.3\, \mathrm{K}$ the inelastic effect displays itself as a plateau
feature appeared at $B=B_{5}$. At lower $T$ it transforms into a wavy line
whose amplitude increases sharply with cooling.

\begin{figure}[tbp]
\begin{center}
\includegraphics[width=11.cm]{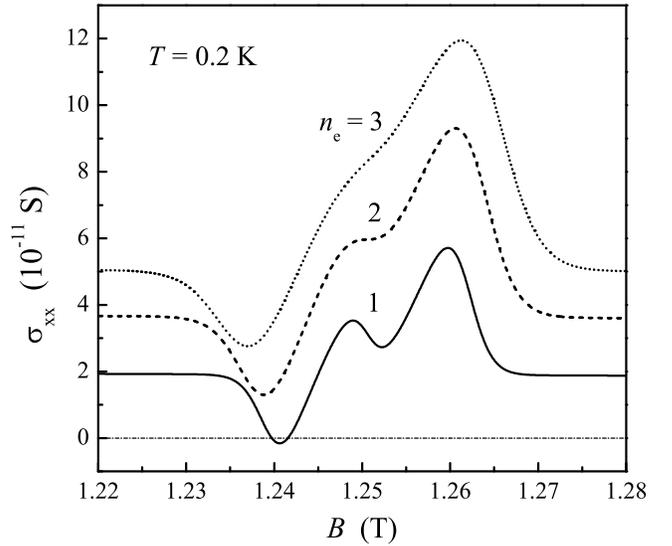}
\end{center}
\caption{Magnetoconductivity vs $B$ near the level-matching point
$B_{4}=1.25\,\mathrm{T}$ for three different electron densities $n_{e}$
shown in units of $10^{6}\,\mathrm{cm^{-2}}$.
} \label{f6}
\end{figure}

The Coulomb broadening of the DSF $\sqrt{x_{q}}\Gamma _{C}$ increases
strongly with electron density. Therefore, with a substantial increase
in $n_{e}$ the inelastic effect becomes
suppressed, as illustrated in Fig.~\ref{f6}.
This figure shows the shape of magnetoconductivity oscillations near $%
B_{4}=1.25\, \mathrm{T}$ for three different electron densities. One can see
that the increase of $n_{e}$ by the factor $3$ eliminates the fine wavy
structure introduced by inelastic scattering.
The Fig.~\ref{f6} can be used also for
modeling of the influence of electron heating on the inelastic effect. Since $%
\Gamma _{C}\propto T_{e}^{1/2}n_{e}^{3/4}$, the increase of $T_{e}$ by the
factor $3^{3/2}$ produces the same Coulomb broadening and the same reduction
of the inelastic effect as the increase of $n_{e}$ by the factor $3$ which
is shown in the figure. It should be noted also that electron heating can affect
the height of the new wavy anomaly by the decrease of
$\nu _{\mathrm{intra}}$ discussed above.

Holding electric field $E_{\perp }$ increases the frequency $\omega _{2,1}$
and the characteristic magnetic fields $B_{m^{\ast }}$ which favours the
inelastic effect. On the other hand, a larger $E_{\perp }$ increases the
collision broadening of Landau levels which reduces the inelastic parameter $%
\hbar \omega _{q}/\Gamma _{l,n}$. Therefore, the holding field should be
kept much smaller than the effective field of the image potential
contributing into the electron-ripplon coupling: $E_{\perp }^{(\mathrm{eff}%
)}\sim 180\, \mathrm{V/cm}$ at $B\sim 1\, \mathrm{T}$. At the same time, in the
limit of low holding fields, there is a chance to meet the resonance
condition for three surface subbands simultaneously, as noted in the previous Section.
In this case, the shape of magnetoconductivity oscillations will be
affected by an additional filling of the third resonant subband $\bar{n}%
_{k}>\bar{n}_{1}e^{-\Delta _{k,1}/T_{e}}$, and there might be additional
variations of $\sigma _{xx}$ at fractional values of the ratio $\omega
/\omega _{c}$ caused by electron scattering from the third subband ($l=k$) to
the ground subband ($l=1$). The results of these evaluations will be given
elsewhere.

\section{Conclusions}

We have presented theoretical description of MW-resonance-induced
magnetoconductivity oscillations for surface electrons on liquid helium
under the condition that the energy exchange at a collision cannot be
neglected as compared to the Landau level broadening. Such a condition can
be realized for SEs on liquid $^{4}\mathrm{He}$ at low temperatures ($T\leq
0.2\, \mathrm{K}$) and weak holding fields. The inelastic effect discussed here
is shown to affect differently the decay rate of the excited subband
and the electron momentum relaxation rate caused by
inter-subband scattering. Near the level matching points ($B\simeq
B_{m^{\ast }}$) the decay maxima are split because of ripplon destruction and
creation processes. This splitting surprisingly becomes stronger in the low
magnetic field range because of the enhancement factor
$m^{\ast }=\mathrm{round}(\omega_{2,1} /\omega_{c})$. At the same time,
magnetoconductivity variations induced by the
inelastic effect represent a mixture of splitting and inversion. As a
result, a new wavy feature can be realized on the shape of magnetoconductivity
oscillations in the vicinity of the
level-matching points. The
amplitude of this fine structure increases in the range of low magnetic
fields, which is contrary to the inelastic effect observed for intra-subband
scattering~\cite{MonItoShi-1997}.

We have studied the influence of strong electron-electron interaction and
possible heating of the electron system on the display of the inelastic
effect in photoconductivity oscillations. The results obtained indicate that
a substantial increase in electron density $n_{e}$ above $1\cdot 10^{6}%
\, \mathrm{cm}^{-2}$ reduces strongly new wavy variations of $\sigma
_{xx}\left( B\right) $ induced by the inelastic effect. Similar reduction in
the display of the inelastic effect is expected if the electron system is
heated strongly because of the decay of excited subbands. Fortunately, the negative
effect of electron heating is caused by the Coulomb broadening of the
generalized dynamic structure factor of the multi-subband 2D electron
system which can be reduced by a proper decrease in electron
density.


\begin{thebibliography}{9}

\bibitem{ZudSim-2001} M.A. Zudov, R.R. Du, J.A. Simmons, and J.R. Reno,
Phys. Rev. B {\textbf{64}}, 201311(R) (2001).

\bibitem{ManSme-2002} R. Mani, J.H. Smet, K. von Klitzing, V. Narayanamurti,
W.B. Johnson, and V. Umansky, Nature \textbf{420}, 646 (2002).

\bibitem{ZudDu-2003} M.A. Zudov, R.R. Du, L.N. Pfeiffer, and K.W. West,
Phys. Rev. Lett. \textbf{90}, 046807 (2003).

\bibitem{DurSac-2003} A.C. Durst, S. Sachdev, N. Read, and S.M. Girvin, Phys. Rev.
Lett. \textbf{91}, 086803 (2003).

\bibitem{DmiVav-2005} I.A. Dmitriev, M.G. Vavilov, I.L. Aleiner, A.D. Mirlin, and
D.G. Polyakov, Phys. Rev. B \textbf{71}, 115316 (2005).

\bibitem{Mik-2011} S.A. Mikhailov, Phys. Rev. B \textbf{83}, 155303 (2011).

\bibitem{KonKon-2009} D. Konstantinov and K. Kono, Phys. Rev. Lett. \textbf{103},
266808 (2009).

\bibitem{KonKon-2010} D. Konstantinov and K. Kono, Phys. Rev. Lett. \textbf{105},
226801 (2010).

\bibitem{ColCoh-1969}  M.W. Cole and M.H. Cohen, Phys. Rev. Lett. \textbf{23},
1238 (1969).

\bibitem{Shi-1970}  V.B. Shikin, Soviet Phys. JETP \textbf{31}, 936. (1970)
[Zh. Eksperim. teor. Fiz. \textbf{58}, 1748 (1970)].

\bibitem{Mon-2011a} Yu.P. Monarkha, Fiz. Nizk. Temp. \textbf{37}, 108 (2011)
[Low Temp. Phys. \textbf{37}, 90 (2011)].

\bibitem{Mon-2011b} Yu.P. Monarkha, Fiz. Nizk. Temp. \textbf{37}, 829 (2011)
[Low Temp. Phys. \textbf{37}, 655 (2011)].

\bibitem{Ryz-1969} V. I. Ryzhii, Fiz. Tverd. Tela \textbf{11}, 2577 (1969)
[Sov. Phys. Solid State \textbf{11}, 2078 (1970)]; V.I. Ryzhii, R.A.
Suris, and B. S. Shchamkhalova, Fiz. Tekh. Poluprovodn. \textbf{20}, 2078 (1986)
[Sov. Phys. Semicond. \textbf{20}, 1299 (1986)].

\bibitem{MonItoShi-1997}  Yu.P. Monarkha, S. Ito, K. Shirahama, and K.
Kono, Phys. Rev. Lett. \textbf{78}, 2445 (1997).

\bibitem{MonKon-book} Yu.P. Monarkha and K. Kono, \textit{Two-Dimensional
Coulomb Liquids and Solids}, Springer-Verlag, Berlin Heildelberg (2004).

\bibitem{KonCheKon-2012} D. Konstantinov, Chepelianskii, and K. Kono, J. Phys. Soc. Jpn.
\textbf{81}, 093601 (2012).

\bibitem{Mon-2012} Yu.P. Monarkha, Low Temp. Phys. \textbf{38}, 451
(2012) [Fiz. Nizk. Temp. \textbf{38}, 579 (2012)].

\bibitem{AndUem-1974} T. Ando and Y. Uemura, J. Phys. Soc. Jpn. \textbf{36}, 959 (1974).

\bibitem{KubMiyHas-1965} R. Kubo, S.J. Miyake, N. Hashitsume, Solid State Phys.
\textbf{17}, 269 (1965).

\bibitem{MonSokSmo-2010} Yu.P. Monarkha, S.S. Sokolov, A.V. Smorodin,
and N. Studart, Low Temp. Phys. \textbf{36}, 565 (2010) [Fiz. Nizk. Temp.,
\textbf{36}, 711 (2010)].

\bibitem{MarLov-1971}  W. Marshall and S.W. Lovesey: \textit{Theory of
Thermal Neutron Scattering} (Clarendon Press, Oxford 1971).

\bibitem{Ger-1976}  R.R. Gerhardts, Surf. Sci. \textbf{58}, 227 (1976).

\bibitem{DykKha-1979} M.I. Dykman and L.S. Khazan, Sov. Phys. JETP \textbf{50},
747 (1979) [Zh. Eksp. Teor. Fiz. \textbf{77}, 1488 (1979)].

\bibitem{KonIssMon-2007} D. Konstantinov, H. Isshiki, Yu. Monarkha, H. Akimoto, K.
Shirahama, and K. Kono, Phys. Rev. Lett. 98, 235302 (2007).

\bibitem{PetSchLea-1994}  P.J.M. Peters, P. Scheuzger, M.J. Lea, Yu.P.
Monarkha, P.K.H. Sommerfeld, and R.W. van der Heijden, Phys. Rev.
B \textbf{50,} 11570 (1994).

\bibitem{MonTesWyd-2002} Yu.P. Monarkha, E. Teske, and P. Wyder,
Phys. Rep. \textbf{370}, No. 1, pp. 1-61 (2002).

\bibitem{AndAle-2003} A.V. Andreev, I.L. Aleiner, and A.J. Millis, Phys.
Rev. Lett., \textbf{91}, 056803 (2003).



\end{thebibliography}
\end{document}